\documentclass[twocolumn,english,aps,prb,showpacs,superscriptaddress,amssymb,amsfonts]{revtex4}
\usepackage[T1]{fontenc}
\usepackage[latin9]{inputenc}
\usepackage{amsmath}
\usepackage{epstopdf}
\usepackage{bm}
\usepackage{graphicx}
\usepackage{amssymb}
\usepackage{color}
\usepackage{enumerate}

\makeatletter
\newcommand{\be}{\begin{equation}}
\newcommand{\ee}{\end{equation}}                  
\newcommand{\bea}{\begin{eqnarray}}
\newcommand{\eea}{\end{eqnarray}}
\newcommand{\beas}{\begin{eqnarray*}}
\newcommand{\eeas}{\end{eqnarray*}}

\newcommand{\avgs}{\overline{S}}
\newcommand{\tr}{\textrm{tr}}
\newcommand{\avgc}{\overline{S}(l,e_c)}
\newcommand{\avgse}{\overline{S}(l,e)}
\newcommand{\avgn}{\overline{S}(l,e^{-}_c)}

\newcommand{\avgng}{\overline{S}(l \gg \xi,e^{-}_c)}
\newcommand{\avgp}{\overline{S}(l,e^{+}_c)}
\newcommand{\avgpl}{\overline{S}(l \ll \xi,e^{+}_c)}
\newcommand{\avgpg}{\overline{S}(l \gg \xi,e^{+}_c)}

\makeatother

\usepackage{babel}

\makeatother

\usepackage{babel}

\begin{document}
 
 \title{Certain General Constraints on the Many-Body Localization Transition}

\author{Tarun Grover}

\affiliation{Kavli Institute for Theoretical Physics, University of California, Santa Barbara, CA 93106, USA}

\begin{abstract}
Isolated quantum systems at strong disorder can display many-body localization (MBL), a remarkable phenomena characterized by an absence of conduction even at finite temperatures. As the ratio of interactions  to disorder is increased, one expects that an MBL phase will eventually undergo a dynamical phase transition to a delocalized phase. Here we constrain the nature of such a transition by exploiting the strong subadditivity of entanglement entropy, as applied to the many-body eigenstates close to the transition in general dimensions. In particular, we show that at a putative continuous transition between an MBL and an ergodic delocalized phase, the critical eigenstates are necessarily thermal, and therefore, the critical entanglement entropy equals the thermal entropy. We also explore a qualitatively different continuous localization-delocalization transition, where the delocalized phase is non-ergodic whose volume law entanglement entropy tends to zero as the transition is approached. 
\end{abstract}

\maketitle

\section{Introduction} \label{sec:intro}
Isolated disordered quantum systems at zero temperature exhibit the phenomena of Anderson localization \cite{anderson1958}: at disorder strong enough compared to kinetic energy and interactions, the ground state is localized. In fact, for a non-interacting isolated disordered system, all many-body eigenstates are localized at any disorder in 1D and 2D, and at strong disorder in 3D. What is the nature of finite-energy density eigenstates in such a system as interactions are turned on? As argued using perturbative methods in Refs.\cite{anderson1980, gefen1997, gornyi2005} and especially Ref.\cite{basko2006}, numerically corroborated in Refs.\cite{huse2007, rigol2007, pal2011,znidari2008,monthus2010,berkelbach2010,gogolin2011,canovi2011,buccheri2011,cuevas2012, bela, luca2013, serbyn2013_1, serbyn2013_2,pollman2012,pekker, pollmann2014} and recently proved in 1D under certain reasonable assumptions \cite{imbrie}, for weak interactions the finite energy density eigenstates continue to remain localized, leading to a ``many-body localized'' (MBL) phase. Thus, for a generic Hamiltonian, one expects that either (i) all states are localized, (ii) all states are delocalized, or (iii) there is a many-body mobility edge separating localized states from delocalized ones. In the case (iii), one therefore undergoes a localization-delocalization transition as a function of the energy density \cite{gefen1997, basko2006}, some of whose properties were studied numerically in Refs.\cite{huse2007,monthus2010, pal2011, pollmann2014}. In this paper, we provide certain general results that constrain the nature of this transition, under the assumption that the transition is continuous. In particular, we show that if the delocalized phase satisfies ``eigenstate thermalization hypothesis'' (ETH) \cite{deutsch1991, srednicki1994,tasaki,rigol}, then the ETH  necessarily  holds right at the transition as well. We also discuss the possibility of a qualitatively different continuous localization-delocalization transition where the delocalized phase does not satisfy ETH.

Throughout, we will borrow the terminology from equilibrium statistical mechanics, especially the notion of a ``phase'' and ``phase transition'', even though thermodynamically, there is no sharp distinction between an MBL phase and a delocalized phase. In the context of this paper, a phase is defined by a set of  eigenstates which (i) have a contiguous energy density in the thermodynamic limit,  (ii) span a finite range of energy density, and most importantly, (iii) share certain properties when the said property is averaged over a tiny energy density window. This definition subsumes the definition of a phase in equilibrium statistical mechanics where equal-time correlator of an appropriate operator generally suffices to distinguish phases from each other. As an example, consider Heisenberg ferromagnetic model in 3D which exhibits a thermodynamics transition at an energy density $e_c$ in the microcanonical ensemble. Here eigenstates with energy density $e < e_c$ have a non-zero magnetization $m$, while those with $e > e_c$ have $m = 0$. In contrast, the localization-delocalization transition leaves its footprints only in dynamical quantities, e.g. conductivity, or non-local objects such as entanglement entropy (EE) \cite{pal2011, bela, pollmann2014}. As an example, the set of eigenstates $\{ \psi\}$, which together constitute an MBL phase, share the property that the EE of almost all eigenstates scale as an area-law, upto logarithmic corrections, in contrast to the volume law scaling in a delocalized phase \cite{pal2011,bela, swingle, defdelocal}.

In this paper we will employ EE of energy eigenstates as an order parameter for the transition between an MBL phase and a delocalized phase, and its singular behavior as a function of the tuning parameter will therefore serve to characterize the phase transition. On that note, let us introduce the notion of an ``ergodic phase''. We define it as a specific class of delocalized phases which satisfy ETH \cite{deutsch1991, srednicki1994,tasaki,rigol}. In particular, in these systems \cite{deutsch1991, srednicki1994,tasaki,rigol}: 

(i) The expectation value and correlators of few-body operators with respect to an eigenstate $\psi$ equal their thermodynamical average i.e. 
\be 
\langle \psi|O|\psi\rangle = \frac{\textrm{tr}\,\, \left(e^{-\beta H} O\right)}{\textrm{tr}\,\,e^{-\beta H}}
\ee
 where $\beta$ is defined so that  $\langle \psi|H|\psi\rangle = \frac{\textrm{tr}\,\, \left(e^{-\beta H} H\right)}{\textrm{tr}\,\,e^{-\beta H}}$.

(ii) The entanglement entropy $S$ for a subregion $A$ with volume $V_A$ ($< V_{\overline{A}}$), corresponding to an eigenstate $\psi$, equals the thermal entropy at a temperature $\beta^{-1}$ defined in (i), i.e.,

\be 
S= {s}_{\textrm{thermal}}(\beta) V_A
\ee
where ${s}_{\textrm{thermal}}$ denotes thermal entropy density. One question of central interest to us is whether there could be a continuous transition between an MBL phase and an ergodic  phase? We will make the notion of a continuous transition explicit in Sec.\ref{sec:mbltoeth} when we discuss this question.  Numerical studies\cite{huse2007, monthus2010, pal2011, pollmann2014} and a recent mean-field analysis \cite{laumann2014} have already suggested the possibility of such a transition though thus far an analytical understanding has been lacking. We now turn to a general inequality satisfied by entanglement entropy, which will be vital  to our discussion.

\section{Constraints on Entanglement Entropy from Strong Subadditivity} \label{sec:ssa}

\begin{figure}
\begin{centering}
\includegraphics[scale=0.6]{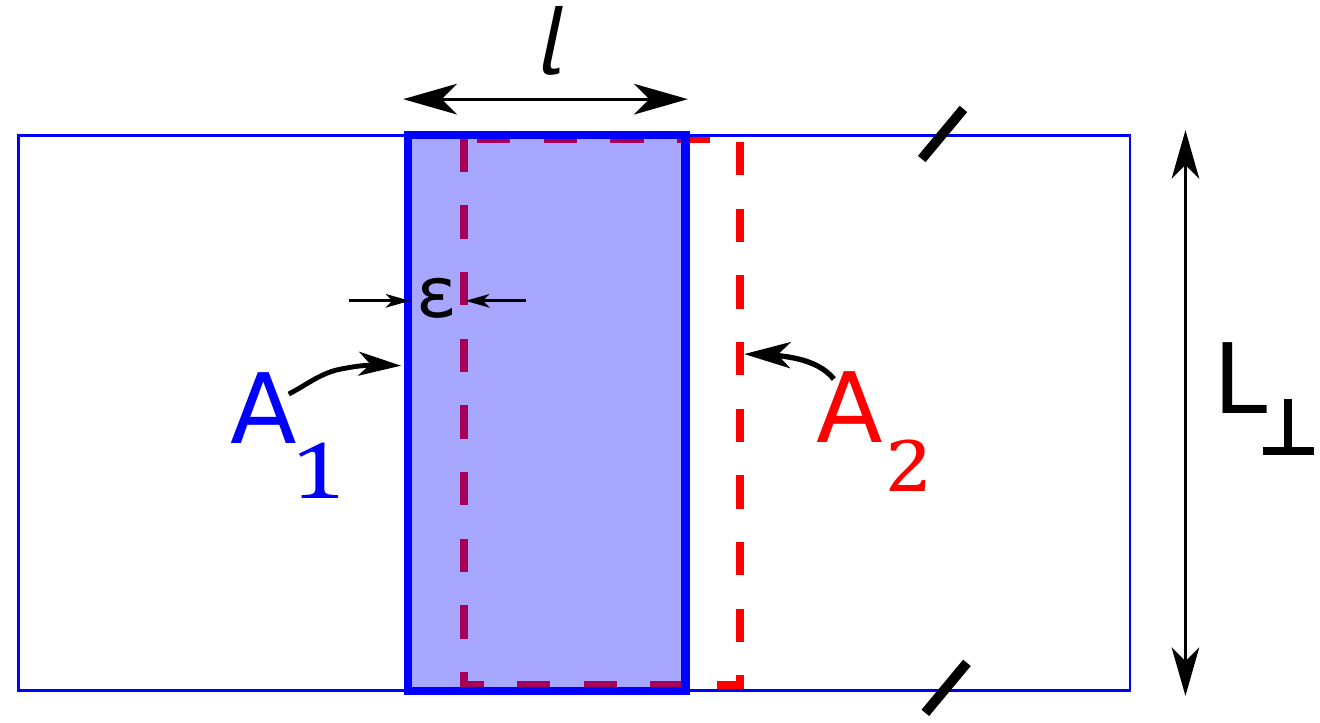}
\par\end{centering}
\caption{The geometry used to derive the concavity constraint $\frac{\partial^2 S(l)}{\partial l^2} \leq 0$ on the entanglement entropy $S(l)$ \cite{hirata2006}. Here region $A_1$ is of dimension $l \times L^{d-1}_{\perp}$ where $L^{d-1}_{\perp} = \prod_{i=2}^{d} L_i $ denotes the size in the perpendicular direction. Region $A_2$ has the same size as $A_1$, and is  displaced relative to $A_1$ by a distance $\epsilon$ ro the right.}\label{fig:bipart}
\end{figure}

\subsection{Concavity of quantum entanglement}
The von Neumann entropy $S$ for an arbitrary density matrix satisfies the strong subadditivity (SSA) inequality \cite{lieb1973}:

\be 
S(A_1) + S(A_2) \geq S(A_1 \cup A_2) + S(A_1 \cap A_2) \label{eq:ssa}
\ee
\noindent where $A_1$ and $A_2$ denote two arbitrary Hilbert spaces. In the context of condensed matter systems, it has been used to derive general results for the RG flow of Lorentz invariant systems, in particular a new derivation of the Zamolodchikov's ``$c$ theorem'' \cite{zamol1986} for 1D Lorentz invariant systems \cite{casini2004}, and it's generalization to 2D, the ``$F$ theorem'' \cite{pufu2011, pufuandothers, casini2012}. For systems that are not Lorentz invariant, SSA is less powerful, but as will discuss below, it still has non-trivial consequences in the context of phase transitions out of an MBL phase.

SSA (Eq.\ref{eq:ssa}) implies that the entanglement is a concave function with respect to appropriate geometric parameters \cite{hirata2006}. To see this, let us consider a $d$ dimensional system of size $\prod_{i=1}^{d} L_i $ with periodic boundary conditions. We bipartition this system into subregions  $A$ and $B$ where $A$ has dimensions $l \times \prod_{i=2}^{d} L_i $ to obtain the reduced density matrix $\rho_A = \tr_B \,\rho $ where $\rho$ is the density matrix corresponding to the total system ($\rho$ could be in a pure or a mixed state). We would be primarily interested in the behavior of $S = -\tr \, (\rho_A \log(\rho_A))$ as a function of $l$, the dimensionality of $A$ along the coordinate axis-1. The application of SSA to the geometry in Fig.\ref{fig:bipart} implies

\be 
2 S(l) \geq S(l+\epsilon) + S(l-\epsilon) \label{eq:ssa2}
\ee
Taking the limit $\epsilon \rightarrow 0$ \cite{footnote:small}, one finds
\be 
\frac{\partial^2 S(l)}{\partial l^2} \leq 0 \label{eq:ineqclean}
\ee

If a system exhibits an RG flow from a UV fixed point to an IR fixed point, one can integrate the above equation to obtain 

\be
\frac{\partial S(l)}{\partial l} \bigg|_{UV} \geq \frac{\partial S(l)}{\partial l} \bigg|_{IR} \label{eq:ineqclean2}
\ee

An additional condition on $S(l)$ is 
\be 
\frac{\partial S(l)}{\partial l} \geq 0,  \label{eq:ineqclean3}
\ee
as long as $l < L_1/2$. This follows from a slightly different formulation of strong subadditivity: $S(A_1 \cup A_2) + S(A_2 \cup A_3) \geq S(A_1) + S(A_3)$. Taking $A_1$ and $A_3$ to be two non-overlapping regions of size $l \times \prod_{i=2}^{d} L_i $, and $A_2$ of size $\epsilon \times \prod_{i=2}^{d} L_i $ such that it is sandwiched between $A_1$ and $A_3$, one recovers the inequality in Eq.\ref{eq:ineqclean3}.

\subsection{Extension to disordered systems} \label{sec:ssadisord}
In the above discussion, we implicitly assumed that the system is translationally invariant -- entanglement $S$ for a subregion $A$ was only a function of the length $l$ and did not depend on the precise location of $A$ within the total system. In a disordered system this is no longer true. The most useful quantity in this context is the disorder averaged EE, which has already  been studied in the context of MBL \cite{pal2011, bela}. To define it precisely, let us diagonalize the underlying Hamiltonian for a fixed disorder realization $\mathcal{D}$, to find the eigenstates $\{\psi_{\mathcal{D}}(e)\}$ where $e$ denotes the energy density eigenvalue corresponding to  a specific eigenstate $\psi_{\mathcal{D}}(e)$. This allows one to obtain the entanglement entropies $\{S_{\mathcal{D}}(l,e)\}$ corresponding to the set $\{\psi_{\mathcal{D}}(e)\}$ for a region $A$ of size $l \times L_2 \times ...\times L_d$. If $\mathcal{D}$ occurs with a probability $P(\mathcal{D})$, we define the disordered averaged entropy $\avgse$ for the region $A$ at energy density $e$ as:

\be 
\avgs(l,e) = \lim_{\Delta e \to 0}  \lim_{\mathcal{V} \to +\infty} \sum_{\mathcal{D}} P(\mathcal{D}) \left(\frac{  \sum_{e'= e - \Delta e/2}^{e' = e+ \Delta e/2}  S_{\mathcal{D}}(l,e')}{\mathcal{N}}\right) \label{eq:defavg}
\ee
where $\mathcal{N}$ is the total number of eigenstates between energy density $ e - \Delta e/2$ and $ e + \Delta e/2 $ and $\mathcal{V}$ is the total volume. Above, we have performed an average over all disorder realizations $\mathcal{D}$ as well as an average over a thin energy shell around $e$. One can now generalize the concavity condition to disordered systems by multiplying the inequality in Eq.\ref{eq:ssa2} for each individual eigenstate that enters the sum in Eq.\ref{eq:defavg} by $P(\mathcal{D}) (\geq 0)$ and summing over all disorder realizations as well as the energy shell. This procedure leads to

\be 
\frac{\partial^2 \avgse}{\partial l^2} \leq 0  \label{eq:ineqdisord}
\ee

One can similarly generalize the inequality in Eq.\ref{eq:ineqclean3} to obtain

\be 
\frac{\partial \avgse}{\partial l} \geq 0,  \label{eq:ineqdisord2}
\ee
for $l < L/2$. In the next section, we will employ Eqns.\ref{eq:ineqdisord} and \ref{eq:ineqdisord2}  to constrain the nature of phase transition out of an MBL phase.
 
\section{Scenarios for transition out of MBL phase}

\begin{figure}
\begin{centering}
\includegraphics[scale=0.5]{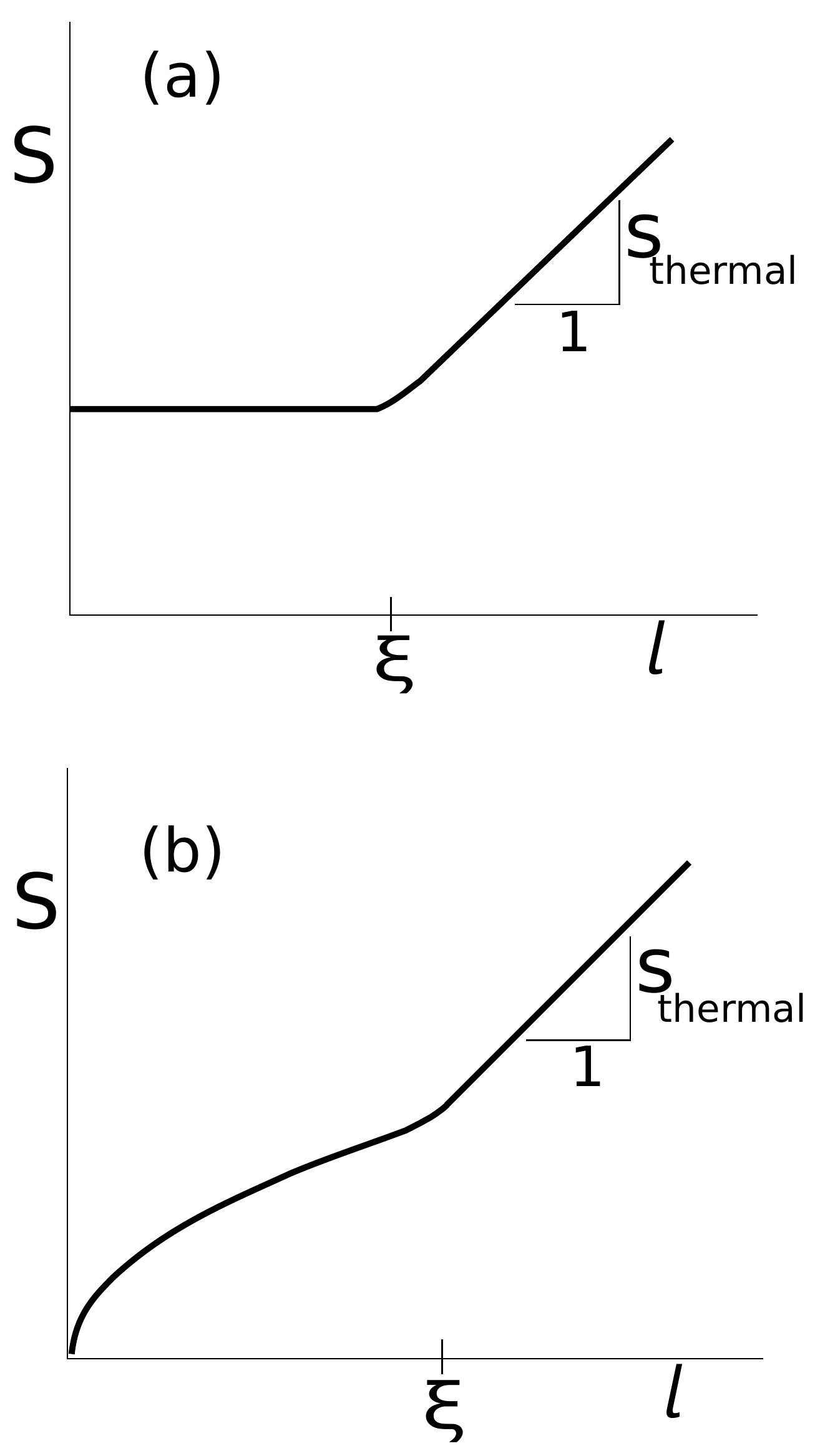}
\par\end{centering}
\caption{Two putative scaling behaviors of entanglement entropy $\overline{S}(l)$ as one approaches the MBL$\leftrightarrow$ ergodic transition from the ergodic side in 1D, both of which are ruled by the concavity condition, Eq.\ref{eq:ineqdisord}. In (a) $S(l) \sim $constant for $l \ll \xi$ and $S(l) \sim l \overline{s}_{\textrm{thermal}}$ for $l \gg \xi$. In (b), $S(l) \sim f(l)$ with $\frac{d^2f}{dl^2}<0$ for $l \ll \xi$  (e.g. $f(l) \sim l^{\alpha}$ or $\log (l)$), and $S(l) \sim l \overline{s}_{\textrm{thermal}}$ for $l \gg \xi$.} \label{fig:eescale}
\end{figure}

\begin{table}
    \begin{tabular}{ | c | c | c | }
    \hline
   \textbf{$\mathbf{\avgs(l)}$ } & \textbf{Ergodic} & \textbf{Non-ergodic} \\ 
      \textbf{at criticality} & \textbf{phase} & \textbf{delocalized phase} \\ \hline
$l^{0} =$ constant & Not allowed & Not allowed \\ \hline
$f(l)$ such that  & Not allowed & Allowed \\ 
$\frac{d^2f}{dl^2} < 0$. (e.g. $\log(l)$)&  &  \\ \hline
$c \,l$  & Allowed only if & Allowed  \\ 
& $c = \overline{s}_{\textrm{thermal}}$ & \\ \hline
\end{tabular}
  \caption{The three possible scenarios for the scaling behavior of the entanglement $\overline{S}(l)$ at a continuous  transition out of an MBL phase to an ergodic, or a non-ergodic delocalized phase in 1D.}
\end{table}

\subsection{Can MBL to ergodic transition be continuous?}  \label{sec:mbltoeth}


Before we proceed, we would like to elaborate  on the notion of a ``continuous phase transition'' in our context. Heuristically, at a continuous transition, the eigenstates evolve continuously from either side of the transition as the transition is approached. More precisely, there exists a length scale $\xi$ on either side of the transition such that on scales $l \ll \xi$, various properties such as the EE of eigenstates or the correlation functions with respect to eigenstates show critical behavior, while for $l \gg \xi$, they approach their value in the respective phase \cite{footnote:onesided}. The continuity of the transition means that the length scale $\xi$ diverges at the transition. For phases that satisfy ETH, this definition matches the conventional definition of a continuous transition in equilibrium statistical mechanics.

We now return to the question posed in the introduction: could an MBL to ergodic  transition be continuous and if yes, what are the possible scaling behavior for the disordered averaged entanglement entropy at the critical point?  We  denote the critical entanglement by $\avgc$, and the entanglement in the MBL and delocalized phases close to the critical point by $\avgn$ and $\avgp$ respectively. All lengths are measured in the units of lattice cutoff $a$, which we set to unity.  For simplicity, we restrict our discussion to 1D in this section and  discuss the generalization to higher dimensions later. Throughout, we will work in the limit $\{L_i\} \gg \xi, l$ so that the finite size effects can be neglected. Let us discuss the results, as summarized in Table I:

(i) $\bm{\avgc \sim}$ \textbf{constant:} Since the EE in an MBL phase satisfies an area law \cite{footnote:log}, this might seem a natural possibility for the scaling of EE in the critical regime as well. This implies that on the ergodic side, $\avgpl = \avgc \sim$ constant, while $\avgpg \sim l \overline{s}_{\textrm{thermal}}(e_c)$ due to ETH, where $\overline{s}_{\textrm{thermal}}(e_c)$ is the disorder averaged thermal entropy density at $e_c$. \textit{However, such a possibility is ruled out by the concavity condition Eq.\ref{eq:ineqdisord}}, as is also obvious from Fig.\ref{fig:eescale}(a). Indeed, on the ergodic side, 
\be 
\frac{\partial \avgs(l)}{\partial l} \bigg|_{UV} = \frac{\partial \avgc}{\partial l} = 0 
\ee
while 

\be 
\frac{\partial \avgs(l)}{\partial l} \bigg|_{IR} = \frac{\partial \avgpg}{\partial l} = \overline{s}_{\textrm{thermal}}
\ee
so that $\frac{\partial S(l)}{\partial l} \bigg|_{UV} < \frac{\partial S(l)}{\partial l} \bigg|_{IR}$, thus violating the concavity condition, Eq.\ref{eq:ineqdisord}.

(ii) $\bm{\avgc  \sim f(l)}$ such that $\bm{\frac{d^2f}{dl^2} < 0 }$:
Two examples of such a behavior are  $f(l) = l^{\alpha}$ with $\alpha < 1$ and $f(l) = \log(l)$, the latter being the most frequently encountered scaling for EE at quantum critical points, including at certain disordered $T=0$ critical points \cite{moore2004}. This implies that on the MBL side, $\avgng = \textrm{constant} \approx f(\xi)$. However, on the ergodic side,  $\frac{d^2f}{dl^2} < 0$  implies that $\frac{\partial \avgc(l)}{\partial l}\bigg|_{l = \xi}$ can be made arbitrary small, if $\xi$ diverges at the transition (Fig.\ref{fig:eescale}(b)). For example, when $f(l) = \log(l)$,
\be 
\frac{\partial \avgc(l)}{\partial l}\bigg|_{l = \xi} \sim \xi^{-1} \geq  \overline{s}_{\textrm{thermal}}(e_c)
\ee
Since $\overline{s}_{\textrm{thermal}}(e_c)$ is of order unity, the divergence of correlation length is incompatible with concavity, Eq.\ref{eq:ineqdisord}. Therefore, this scenario is also ruled out.
%
%

\begin{figure}
\begin{centering}
\includegraphics[scale=0.6]{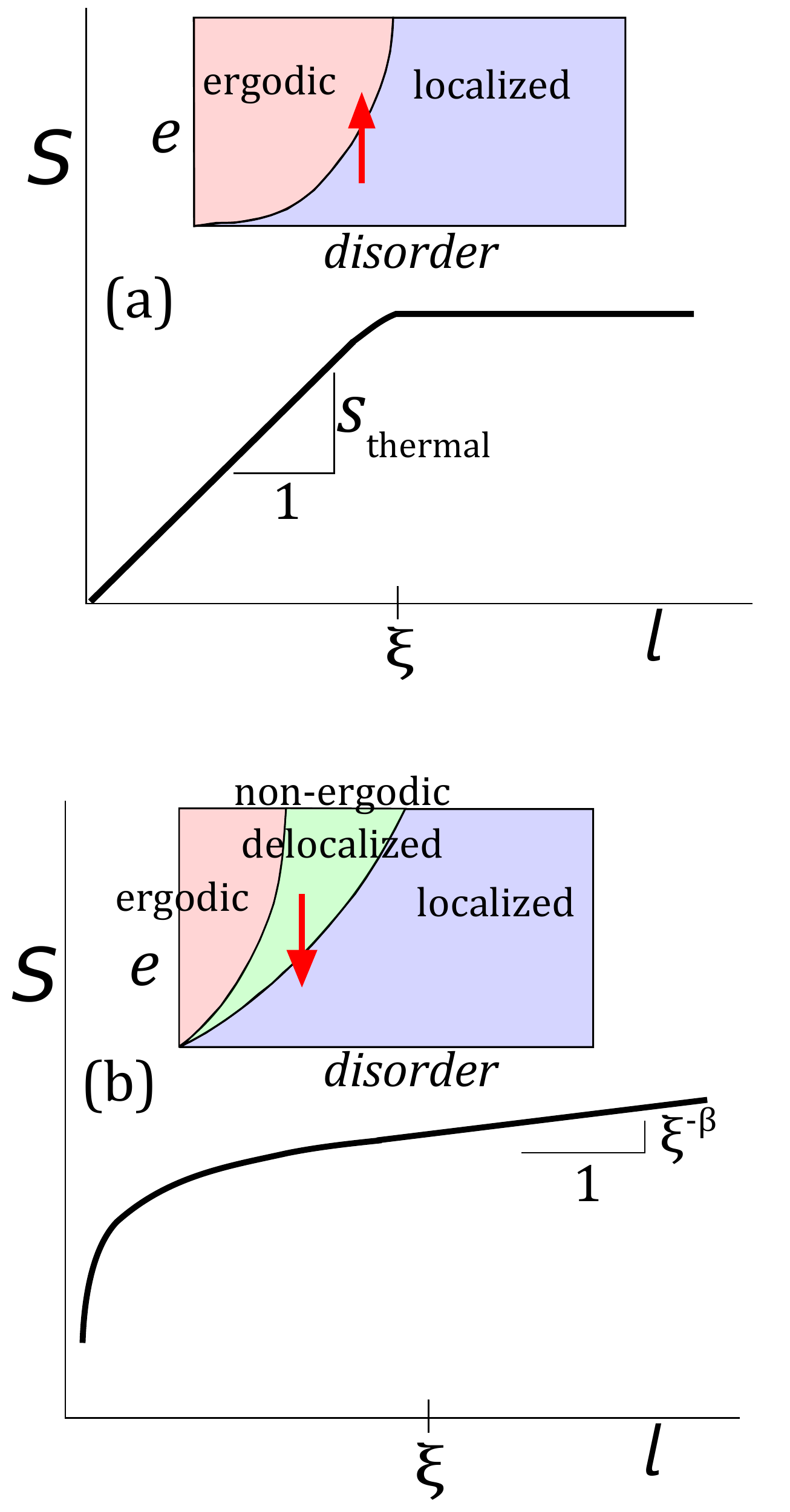}
\par\end{centering}
\caption{Two qualitatively different scenarios for a finite energy density localization-delocalization transition in 1D, and the corresponding two schematic phase diagrams in the disorder-energy density ($=e$) plane. In (a),  the transition is approached from the localized side and the delocalized phase satisfies ETH. In (b), the transition is instead approached from a non-ergodic delocalized phase. In (b), $S$ is sub-volume law for $l \ll \xi$, and volume-law for $l \gg \xi$. $\beta$ is determined by demanding continuity of $\overline{S}$ and $\frac{\partial \overline{S}}{ \partial l}$ (for example, if $S(l \ll \xi) \sim \log(l)$, then $\beta = 1$). } \label{fig:eescale2}
\end{figure}

(iii) $\bm{\avgc  \sim l}$: As implied by (i),(ii) above, this is the \textit{only} scenario potentially allowed for a continuous transition between an MBL phase and an ergodic phase. Concavity implies that $\avgc \geq \overline{s}_{\textrm{thermal}} l$. On the other hand, $\avgc$ also satisfies an upper bound $\avgc \leq \overline{s}_{\textrm{thermal}} l$ which follows from the positivity of relative entropy and is saturated when the reduced density matrix is thermal (Appendix A). Therefore, 
\be \avgc = \overline{s}_{\textrm{thermal}}(e_c) l
\ee 
It is remarkable that the concavity condition (Eq.\ref{eq:ineqdisord})  is sufficiently constraining to deduce the nature of eigenstates at the transition. Thus, as one approaches the transition from the MBL side, for $l \ll \xi$, the system is ergodic while for $l \gg \xi$, $S(l)$ saturates to a constant of order $\overline{s}_{\textrm{thermal}}(e_c) \xi$ (Fig.\ref{fig:eescale2}(a)). At the transition itself, the system is fully thermalized (i.e., satisfies ETH). Note that in this scenario, system manages to exhibit a continuous transition even though there is no cross-over when $\xi$ becomes comparable to $l$ on the ergodic side of the transition.  

One might wonder that in the above scenario, even though at the leading order the critical entanglement satisfies ETH, might it differ from it at the subleading order? The most natural candidate for a universal subleading correction to $\avgc$ is a constant term since it does not necessarily involve any short distance physics (such as lattice constant $a$ or $\overline{s}_{\textrm{thermal}}(e_c)$). Interestingly, even this possibility is ruled out by the concavity condition. To see this, we first note the constant term should come with a negative sign, i.e., $\avgpl = \overline{s}_{\textrm{thermal}}(e_c) l - \gamma$ (with $\gamma > 0$), since $\overline{s}_{\textrm{thermal}} l$ saturates the EE upper bound (Appendix A). At long distances, this must asymptote to $\avgpg = \overline{s}_{\textrm{thermal}}(e_c) l$. However, as may be readily verified, such as function $\overline{S}(l,e_c^{+})$ violates concavity.

Finally, we ask whether the delocalization transition out of MBL, irrespective of whether the delocalized phase satisfies ETH or not, could be \textit{first-order} in nature? At such a transition, the nature of many-body eigenstates changes abruptly from localized to delocalized at a certain  energy density $e_c$, without any diverging length scale from either side of the transition. Generically, this means that the function $ \avgse$ will be discontinuous across a critical energy density $e_c$ for almost all fixed $l$. Though we can't rule out such a transition \cite{footnote:mott}, we do not know of a similar transition even within equilibrium statistical mechanics, which seems to suggest that it might be very unlikely \cite{footnote:firstorder}.

\subsection{Possibility of a non-ergodic delocalized phase}  \label{sec:noneth}
The discussion in the previous  section motivates a completely different kind of delocalization transition via an intermediate non-ergodic delocalized phase. Here, as one approaches the transition from the delocalized side, the coefficient of the volume law vanishes, e.g., $\avgp \sim l/\xi^{\beta}$ ($0 < \beta \leq 1$) where $\xi$ is the correlation length, thus smoothly connecting to an area law entanglement on the localized side (Fig.\ref{fig:eescale2}(b)) \cite{footnote:subvolume}. Therefore, ETH is violated in the delocalized phase as well. Note that unlike the previously discussed continuous transition between MBL and ergodic phase, here the entanglement $S(l,e)$ shows a crossover on both sides of the transition when $l$ passes through $\xi$. Let us ask two basic questions about such a non-ergodic phase:

(i) Can a non-ergodic delocalized phase be connected to an ergodic one without a phase transition?

(ii) Are there any restrictions on the form of entanglement scaling at the critical point between the localized and the non-ergodic delocalized phase?

The answer to the first question is in the negative:  the function $f(e) = S(e) - \overline{s}_{\textrm{thermal}}(e)$ vanishes in the ergodic phase and is non-zero in the non-ergodic phase as one tunes $e$ across the ergodic to non-ergodic transition. Therefore, $f(e)$ must have a singularity at an intermediate energy density \cite{footnote:ergodic}. Of course, as before, we do assume that the eigenstates that belong to ergodic (or non-ergodic) phase are contiguous in energy so that the notion of phase is well-defined.

Regarding the question (ii), the results are summarized in Table I. Essentially, the only constraint as one approaches from the delocalized side is that $\avgc$ cannot be a strict area-law since it will violate concavity.

Finally, we note that at a putative continuous transition between the non-ergodic delocalized phase and the ergodic phase (Fig.\ref{fig:eescale2}(b)), the EE will satisfy ETH, via the same reasoning as in Sec.\ref{sec:mbltoeth}.

\subsection{Generalization to Higher Dimensions} \label{sec:higherd}
Since the inequality in Eq.\ref{eq:ineqdisord} was derived in general dimensions, all results discussed in Sections \ref{sec:mbltoeth} and \ref{sec:noneth} for 1D continue to hold in higher dimensions with the replacement $\overline{S} \rightarrow \overline{S}/L^{d-1}_\perp$. In particular, at an MBL to ergodic transition, the system must again be thermalized even at the critical point and the critical entanglement entropy for a subregion of volume $V$ equals
\be
\overline{S} = \overline{s}_{\textrm{thermal}}(e_c)V
\ee

\section{Discussion} \label{sec:discuss}
In this paper, we employed the strong subadditivity (SSA) inequality of quantum entanglement to put strong constraints on the nature of a continuous localization-delocalization transition at finite energy densities. In particular, we showed that at a transition between an MBL phase and an ergodic phase, the critical eigenstates satisfy ETH in all dimensions (Fig.\ref{fig:eescale2}(a)). The constraints due to SSA also lead us to explore a completely different kind of localization-delocalization transition where the delocalized phase does not satisfy ETH (Fig.\ref{fig:eescale2}(b)). In fact, if one naively considers the coefficient of volume law term in the delocalized phase as the order-parameter for the localization-delocalization transition, then one would reach the erroneous conclusion that a continuous transition out of MBL always falls under the latter scenario.  Which of these two scenarios is realized in a physical system, such as the model Hamiltonians studied numerically in Refs.\cite{huse2007, rigol2007, pal2011,znidari2008,monthus2010,berkelbach2010,gogolin2011,canovi2011,buccheri2011,cuevas2012, bela, luca2013, serbyn2013_1, serbyn2013_2,pollman2012}? As of now the only tool available to investigate MBL transition is exact diagonalization (ED) which is limited to very small sizes ($L \lesssim 20$). Such small systems make it  difficult to access the true nature of the critical point. However, Refs.\cite{huse2007, pal2011}, based on  ED numerics suggested that the behavior at the critical point may be more like a localized phase than like an ergodic phase.  They in fact suggested that it might be an infinite-randomness fixed point. If the localizing character of the critical eigenstates seen in dynamics carries over to the scaling of entanglement as well, then it follows from our discussion that the delocalized phase is non-ergodic. However, the ability of thermalize does not imply that the system is necessarily conducting since the thermalization time scales will typically be much larger than the time to diffuse across the system \cite{footnote:thermal}. Furthermore, Ref.\cite{pollmann2014} provided numerical support for a continuous transition between MBL and ergodic phase by scaling analysis of the entanglement entropy generated due to local quenches. In their work, the EE at the critical point for an equal bipartition of the total system scales linearly with the total system size to a good approximation (i.e., a volume law), and is subthermal. This might be consistent with the ETH being satisfied at the critical point, once the finite size effects are taken into account \cite{husediscuss}.

It is interesting to contemplate the possibility of the non-ergodic delocalized phase discussed in this paper, even if the model Hamiltonians studied in Refs.\cite{huse2007, rigol2007, pal2011,znidari2008,monthus2010,berkelbach2010,gogolin2011,canovi2011,buccheri2011,cuevas2012, bela, luca2013, serbyn2013_1, serbyn2013_2,pollman2012,pekker, pollmann2014} happen to not support such a phase. Even though such a phase in itself would be rather exotic, it is amusing that a continuous phase transition between such a phase and a localized phase would be virtually more conventional, compared to the one between an ergodic and a localized phase, as explained in Sec.\ref{sec:noneth}. We also note that a delocalized non-ergodic phase is reminiscent of ``soft chaos'' generally discussed in the context of KAM theorem as applied to the classical systems with finite degrees of freedom \cite{arnold}. This is because an MBL phase is effectively integrable due to an extensive number of local conserved quantities \cite{huse2013,serbyn2013_2} and a continuous delocalization transition is the analog of breaking integrability slightly. Therefore, the question whether an infinitesimal breaking of integrability leads to only partial thermalization, is intimately tied with the possibility of the existence of such a phase. Numerical work on a certain class of integrable quantum many-body systems  \cite{rigol_kam} seems to suggest that an infinitesimal perturbation is sufficient to restore ``quantum chaos'' for almost all states but we are unaware of any rigorous results in this direction. Can one find model Hamiltonians where the localization-delocalization transition in Fig.\ref{fig:eescale2}(b)  is realized? Does a non-ergodic delocalized phase, if it exists, necessarily conduct? What is the nature of eigenstates that can lead to sub-thermal volume law entanglement entropy? We leave these intriguing questions to the future.

\textit{Acknowledgements:} I thank David Huse, Leon Balents, and Matthew Fisher for stimulating discussions and helpful comments on the manuscript, and Nabil Iqbal for helpful discussions.


\appendix

\section{Upper bound on EE: $s \leq s_{\textrm{thermal}}$}

Let us consider a Hamiltonian $H$ with eigenstates $\{\psi(e)\}$, and thermal entropy density $s_{\textrm{thermal}}(e)$ where $e$ is the energy density corresponding to $\psi$. Here we show that for a given bipartition of the total system into subregions $A$ and $\overline{A}$, the entanglement entropy density corresponding to an eigenstate with entropy density $e$ is bounded from above by $s_{\textrm{thermal}}(e)$.

Denoting the projection of $H$ onto region $A$ as $H_A$, consider an auxiliary density matrix $\sigma(\beta)$ with support on $A$ defined as:

\be 
\sigma(\beta) =  \frac{e^{-\beta H_A}}{\textrm{tr}\,\,e^{-\beta H_A}}
\ee
where $\beta$ is a free parameter. We will employ the following inequality \cite{nielsen} that holds for two arbitrary density matrices $\rho_1, \rho_2$:

\be 
\tr\left(\rho_1 \log \rho_1\right) - \tr\left(\rho_1 \log \rho_2\right) \geq 0
\ee
where the equality holds if and only if $\rho = \sigma$. Taking $\rho_1$ as the reduced density matrix corresponding to $\psi(e)$ for the bipartition $A, \overline{A}$ and $\rho_2 = \sigma$, the above inequality implies

\be 
S(\rho_1) = -\tr\left(\rho_1 \log \rho_1 \right) \leq \widehat{S}_{\textrm{thermal}}(\beta) \label{eq:boundthermal}
\ee
where $\widehat{S}_{\textrm{thermal}}(\beta) = \beta(E-F(\beta))$, $E = \tr \left( \rho H \right) = e V_A$ and $\beta F = -\log( \tr\, e^{-\beta H_A})$. Note that $E$ is independent of $\beta$ and $V_A$ is the volume of region $A$. Chosing $\beta$ such that $\tr\, \left(\sigma(\beta) H_A\right) = E$, $ $ Eq.\ref{eq:boundthermal} implies 
\be 
S(\rho_1) \leq s_{\textrm{thermal}}(\beta) V_A = s_{\textrm{thermal}}(e) V_A  \label{eq:boundthermal2}
\ee
where we have used the equality between thermal entropies in the canonical and microcanonical ensemble. Furthermore, one may show that $s_{\textrm{thermal}}(\beta) V_A $ is the best upper bound implied by Eq.\ref{eq:boundthermal} on $S(\rho_1)$. This follows by minimizing $\widehat{S}_{\textrm{thermal}}(\beta)$ with respect to $\beta$:

\be 
d\widehat{S}_{\textrm{thermal}}(\beta) = (E - E(\beta))\, d\beta
\ee
where $E(\beta) = \tr\, \left(\sigma(\beta) H_A\right)$. Thus $\widehat{S}_{\textrm{thermal}}(\beta)$ is extremized when $\beta$ satisfies $E = E(\beta)$. It is easy to see this corresponds to global minima by evaluating $\widehat{S}_{\textrm{thermal}}(\beta)$ at the endpoints $\beta = 0, \infty$. When $\beta \rightarrow 0$, $\widehat{S}_{\textrm{thermal}}(\beta) \rightarrow s_{\textrm{thermal}}(\beta = 0)V_A$, which is the global maximum of $s_{\textrm{thermal}}(\beta = 0)V_A$, while as $\beta \rightarrow \infty$, $\widehat{S}_{\textrm{thermal}}(\beta)$ diverges.

\textit{Generalization to Disordered Systems:} By disorder averaging both sides of the  inequality in Eq.\ref{eq:boundthermal2} via the definition in Eq.\ref{eq:defavg}, one obtains $\overline{S} \leq V_A \overline{s}_{\textrm{thermal}}$.


\begin{thebibliography}{x}
\bibitem{footnote:small} In a system regularized on a lattice with lattice cutoff $a$ and a correlation length $\xi \gg a$, the limit $\epsilon \rightarrow 0$ should me interpreted as $\epsilon/d \rightarrow 0$ where $d = \textrm{minimum}(l, \xi)$, while $\epsilon \gtrsim a$.
\bibitem{footnote:onesided} As discussed below, there exist exceptions where the correlation length diverges only from one side of the transition while it  still being a continuous transition.
\bibitem{footnote:log} For specific Hamiltonians, see e.g. Ref.\cite{pekker}, the area law could be violated by multiplicative logarithms. This does not modify any of our conclusions and for simplicity, we will assume that MBL eigenstates satisfy a strict area law.
\bibitem{footnote:firstorder} We recall a first-order transition in equilibrium statistical mechanics occurs when $\frac{\partial^2 s}{\partial e^2}$ changes its sign, while the function $s(e)$ remains continuous.
\bibitem{footnote:subvolume} As a rather special case, a volume law entanglement at the criticality is also allowed, as long as it does not satisfy ETH (see third column, Table I).
\bibitem{footnote:thermal} For example, a time-evolved product state will resemble a thermal state (from the standpoint of ETH), only after time-scale of order inverse many-body level spacing which scales as $e^{L^d}$. The time to diffuse across the system on the other hand scales only as $L^2$.
\end{thebibliography}
\end{document}